# Climate Control Using Nuclear Energy

**Author: Moninder Singh Modgil**
**Email: moni_g4@yahoo.com**

**Abstract:**
We examine implications of anthropogenic low pressure regions, - created by injecting heat from nuclear reactors, into atmosphere. We suggest the possibility that such artificially generated low pressure regions, near hurricanes could disrupt their growth, path, and intensity. This method can also create controlled tropical stroms, which lead to substantial rainfall in arid areas, such as - (1)Sahara desert, (2) Australian interior desert, and (3) Indian Thar desert. A simple vortex suction model is developed to study, effect on atmospheric dynamics, by such a nuclear heat injection system.

# 1. Background

## 1.1 Anthropogenic Low Pressure Regions

Amospheric codes, can be used to study effect of low pressure regions, created using injection of heat from nuclear reactor, into atmosphere, on circulation over Oceans and in the Ocean-land systems. Such scenarios could arise by accident – such as a run away nuclear reactor of a ship or a submarine in ocean, or a coastal nuclear power plant. It could also be intentional - an example of geo-engineering – where nuclear energy is used to control and/or reduce intensity of hurricanes or even create rainfall in deserts, by pulling in moist ocean air, into interior of the desert. Some more conventional examples, of man made low pressures regions, are large coastal cities such as New York, Bombay - where, the low pressure region is caused by heat injected into atmosphere by electricity consumption, vehicular emissions, thermal power plants, etc.. Another interesting example, is the Kuwait oil well fires, started by Iraqi soldiers, during pull out from Iraq, in the last Gulf war. The heat injected into the atmosphere by the fires, temporarily modified local weather, and lead to severe sustained rainfall. Evidently, the low pressure region caused by heat input from fires, drew in moist air from Arab sea, and aerosols release by fires provided condensation nuclei. Such scenarios also occur in large forest fires near coastal areas. The focus of this research proposal is however, when nuclear energy becomes the source, of the heat injection, into the atmosphere.

## 1.2 Damage caused by Hurricanes

Katrina, a category 5 hurricane, was the costliest is US history, and inflicted an economic damage of nearly US $ 200 billion, aside from the incalculable human loss. Tropical cyclones on the open sea cause large waves, heavy rain, and high winds, disrupting international shipping and sometimes sinking ships. However, the most devastating effects of a tropical cyclone occur when they cross coastlines, making landfall. A tropical cyclone moving over land can do direct damage in four ways:

- High winds - Hurricane strength winds can damage or destroy vehicles, buildings, bridges, etc. High winds also turn loose debris into flying projectiles, making the outdoor environment even more dangerous.
- Storm surge - Tropical cyclones cause an increase in sea level, which can flood coastal communities. This is the worst effect, as historically cyclones claimed 80% of their victims when they first strike shore.
- Heavy rain - The thunderstrom activity in a tropical cyclone causes intense rainfall. Rivers and streams flood, roads become impassable, and landslides can occur. Inland areas are particularly vulnerable to freshwater flooding, due to residents not preparing adequately.
- Tornado activity - The broad rotation of a hurricane often spawns tornadoes. Also, tornadoes can be spawned as a result of eyewall mesovortices which perisist until landfall. While these tornadoes are normally not as strong as their non-tropical counterparts, they can still cause tremendous damage.

**Alarming global warming links** between apparent increase in hurricane intensity and frequency, have been reported (Webster et al., 2005, Emanuel 2005), which however, are disputed (Gray 2006a, Gray 2006b).

There also exists possibility of **increase in frequency and intensity**, in a specific time period, merely as a *statistical fluctuation*.

### 1.3 A Geo-Engineering Precedent for Hurricane Control

Using geo-engineering approach to mitigating hurricane damage, has an attempted precedent. In the 1960s and 1970s, the United States government attempted to weaken hurricanes in its "*Project Stromfury*" by seeding selected storms with silver iodide. It was thought that the seeding would cause supercooled water in the outer rainbands to freeze, causing the inner eyewall to collapse and thus reducing the winds. The winds of Hurricane Debbie dropped as much as 30 percent, but then regained their strength after each of two seeding forays. In an earlier episode, disaster struck when a hurricane east of Jacksonville, Florida, was seeded, promptly changed its course, and smashed into Savannah, Georgia. Because there was so much uncertainty about the behavior of these storms, the federal government would not approve seeding operations unless the hurricane had a less than 10 percent chance of making landfall within 48 hours. The project was dropped after it was discovered that eyewall replacement cycles occur naturally in strong hurricanes, casting doubt on the result of the earlier attempts. Today it is known that silver iodide seeding is not likely to have an effect because the amount of supercooled water in the rainbands of a tropical cyclone is too low

### 1.4 Deserts and Humans

Deserts cover one third of land surface and inhabit one sixth of human population (Kormondy, 1996). Burgeoning even of an exponentially growing human population, has lead to severe depletion of earth's forest cover, as well as extinction and near extinction of a large number of plant and animal species. This has also altered regional and global weather patterns. Given likelihood of further increases in human population, it makes sense to adopt geo-engineering approach for a large scale transformation of deserts. Purpose of such a transformation would be - (1) To increase forest cover, (2) Agricultural purposes, and (3) Increase hospitable land available for human habitation.

### 1.5 Hurricane Energetics

An order of magnitude estimate of kinetic energy of winds in the severest category 5 hurricane, can be arrived at as follows. Consider a control volume of air, with a circular horizontal area of a million square kilometers, and a height of ten kilometers. This horizontal area corresponds in magnitude, to a circular disk of radius 1000 kilometers. The maximum wind speeds in a category 5 hurricane, are about 300 Kilometers per hour. Let us assume that the average horizontal wind speed, in this control volume is 100 kilometers per hour. This gives an order of magnitude of kinetic energy of hurricane as $10^{18}$ Joules. The storm intensifies due to certain energy feed back mechanisms, and reaches this intensity, over a period of about 10 days. The magnitude of energy of 10, one Giga Watt reactors, operating for a period of 10 days is $10^{16}$ Joules. This figure is two orders of magnitude smaller, than energy of a category 5 hurricane, but is comparable to energy of a category 1 hurricane. _However, the energy feed back mechanisms operating in hurricane will also be available to ANHISHA_. It makes better sense to attempt control, in nascent stages of hurricane. It is possible that, only a much smaller energy input from the reactors may be required to disrupt the hurricane dynamics over mid-ocean, or change its trajectory.

**1.6 Desert Atmospheric Energetics**

Now consider the second application of ANHISHA – namely, creation of rain over deserts. Peak solar energy flux incident on earth's surface, at noon time on equator during equinox, is 1.35 Kilo Watts. Averaging over a 24 hour period, and accounting for high latitude for deserts such as those of southern US, and Australian interior, brings down this figure, by an order of magnitude. Thus, an order of magnitude estimate of daily average solar energy flux, over an area of a million square kilometers, is $10^{14}$ Watts. Sand absorbs only twenty percent incident radiation. Further, the value of heat transfer coefficient between sand and air is very small. Assuming that only one thousand-th of energy absorbed by sand is transferred to the atmosphere, the energy flux into the atmosphere over an area of a million square kilometers is equal to power output of 10, one Giga Watt nuclear reactors. _Therefore, the desert atmospheric energetics, are comparable to nuclear reactor energetics._

**1.7 Global Warming Issue**

Note that the method proposed here is not expected to contribute significantly to global warming, because the prime reason for global warming is increased concentration of green house gases, which however are not released in ANHISHA scenarios. The heat generated by ANHISHA nuclear reactors, is expected to eventually escape from atmosphere into space - analogous to the way, that the heat injected by sun during the day, eventually escapes (mostly during the same night), and essentially maintaining, earth's heat balance. Any additional warming would only be due to pre-existing green house gases.

**1.8 Safety Issue**

Safety issues related to nuclear energy appear to have been sorted out by experts working in this area, as is evident from operation of a large number of nuclear propelled ships, nuclear submarines, and nuclear power stations. Environment in immediate vicinity of ANHISHA's desert and hurricane scenario establishments, would be storm conditions. Personnel and equipment safety would have to be appropriately planned in their design. Safety of nuclear ships being used in ANHISHA's hurricane scenario, would require study, to determine safe distance vectors, from hurricane eye to ship's location, to ensure ship safety..

# 2. Objectives

Earth's atmosphere is indeed driven by nuclear energy! Sun is the driving force behind earth's weather, and is fuelled by nuclear (fusion) energy. Nuclear fission energy drives a number of power stations, ships and submarines. It would be almost logical to use nuclear energy to combat natural phenomena such as hurricanes, and drought prone deserts. A controlled, slow injection of heat energy - equivalent in magnitude, to that of a nuclear bomb, over a period of few days, from a nuclear reactor into atmosphere, over ocean, will create a large and sustained, low pressure region, which could significantly effect local wind circulation and weather patterns. By creating such artificial low pressure regions, using mobile, ship based, nuclear reactors, near hurricanes, it may be possible to control their intensity and trajectories, and mitigate large scale damage  - which for instance, occurred in the Katrina Hurricane (see figure 1).  Aim of the computational study is to examine these possibilities.

Another envisaged application is creating desert rain. Nuclear heat injection systems for heating atmosphere could be used in deserts to pull in moist air from near by oceans and induce rainfall (see figure 2).

This _Application of Nuclear Heat Injection Systems for Heating Atmosphere_ will be given the acronym **ANHISHA**

Let the physical fields, such as - velocity, pressure, and humidity, in a control volume $V$, be denote by $v$, $p$, and $h$, respectively. In general, $v$, $p$, and $h$, will  be functions of spatial coordinates $r$, and time $t$. Our objective is to understand,  time evolution of $v$, $p$, and $h$ when ANHISHA's heat sources are introduced into the local atmosphere We are interested in, effect of ANHISHA's perturbations on – (1) hurricane evolution, and (2) humidity levels and precipitation, in desert interior. For this purpose we wish to simulate ANHISHA scenarios using atmospheric codes. Introduce $N$, heat sources of power $P_i$ , (where, the index $i \; \square \; \{1,2,...N\}$ ),   at position vectors $X_i$ ,into the control volume $V$. We wish to study how the physical fields vary in response to variation in the parameters, $N$, $P_i$, and $X_i$. Accordingly, we propose to run the atmospheric codes for a number of combinations of $N$, $P_i$, and $X_i$ . Interpolation  for $P_i$ and $X_i$,  can be used for calculating the values of the physical fields, at points on which the code has not been run. In particular, we wish to see what combination of  $N$, $P_i$, and $X_i$, gives the desired behavior of physical fields. Desired behavior for hurricane scenario would be hurricane velocity field, $v{\rightarrow}0$, a disruption of feed back mechanism, or a desired change in hurricane path. The desired change in hurricane path, would be to –(1) path modification, e.g.,prevent hurricane from moving over warm water, where its intensity would increase, due to the triggering of the feed back mechanisms, (2) disrupting the hurricane by wind shear. Desired behavior for desert rain scenario, would be increase in humidity field $h$, in desert interior, with relative humidity close to 100.

# 3. Technical Plan

Below we give qualitative arguments, on how the twin objectives of – (1) mitigating hurricane damage, and (2) creating desert rain, can be achieved by ANHISHA.

## 3.1 The Hurricane Control Objective

## 3.1.2 Hurricane Path Modification using ANHISHA

When a tropical cyclone moves into higher latitude, its general track around a high-pressure area can be deflected significantly by winds moving toward a low-pressure area. Such a track direction change is termed *recurve.* A hurricane moving from the Atlantic toward the Gulf of Mexico, for example, will recurve to the north and then northeast if it encounters winds blowing northeastward toward a low-pressure system passing over North America. Many tropical cyclones along the East Coast and in the Gulf of Mexico are eventually forced toward the northeast by low-pressure areas which move from west to east over North America. Low pressure regions created by ANHISHA could therefore modify hurricane paths. One could for instance, draw the hurricane away from warm waters, and there by avoid increase in its intensity. One could in principle, avoid hitting heavily populated coastal areas.

### 3.1.3 Disrupting Hurricanes using ANHISHA

When wind shear is high, the convection in a cyclone or disturbance will be disrupted, blowing the system apart. ANHISHA's suction would generate horizontal winds, directed towards the nuclear heat source, and a large thermal upcurrent. An optimization of location of heat source and its power, could sniff out a hurricane in nascent stages of development.

### 3.2 The Desert Rain Objective

A natural low pressure region, forms over desert during day time, due to solar heating. This is replaced by a high pressure region, during night, as the heat absorbed during day time, is quickly lost by the sandy desert surface. Thus, the air above desert goes through dirnal oscillations in pressure. While during the day, there is tendency of moist air over a neighboring ocean, to flow into desert, this is however reversed during night. The ensuing, oscillating direction of diurnal ocean-desert circulation, is therefore not able to push moist ocean air, into desert interior. ANHISHA with its constant operation, or operation during night, can maintain the low pressure region, continuously. The constant suction generated by ANHISHA would pull moist ocean air, deep into desert interior. When sufficiently high humidity levels have been attained in the desert atmosphere, the nightly cooling of desert air, is expected to induce precipitation.

### 3.2 Vortex-Suction Model for ANHISHA

We have developed a simple, analytical vortex-suction model for ANHISHA, to relate nuclear power with the induced atmospheric velocity field. Various geometric parameters of the vortex-suction model are indicated in figure 3. We use a cylindrical coordinate system $(r,z,\varnothing)$ , where $z$, is the vertical direction (positive upwards), ANHISHA's control volume is a cylinder of radius $R$, and height $H$. Origin $O$, of the co-ordinate system, is also the location of nuclear reactor. Suction effect of nuclear reactor, is felt out to the distance $R$, which can be regarded as the distance, where the induced radial inward velocity is 10 Kilometers per hour – which corresponds to a mild breeze. $R$ will be termed effective suction radius. $H$ is expected to be about 1 Kilometer. The suction generated by injection of nuclear energy is simulated by a vertical line-sink of strength $\Lambda$, and circulation is simulated by a vertical line-vortex of strength $\Gamma$ - both placed at the origin. The thermal up-current is assumed to be confined to a vertical cylinder of radius $r_c$, and will be referred to as the vortex core. ANHISHA's heat is injected into atmosphere, within the vortex core. Radial and tangential velocities, within the vortex core are taken to be zero.

### 3.2.1 Inviscid Energetics

We assume that no energy feed back mechanisms have come into play, during the initial phase of operation of nuclear reactor. The suction and circulation increase with time of operation $\tau$ of the nuclear reactor, i.e., kinetic energy of atmospheric velocity fields, increases, as the nuclear reactor, continues to operate. Induced radial, tangential and vertical velocities are labeled $v_r$, $v_t$, and $v_z$ respectively, and the corresponding kinetic energy components are labeled $E_r$, $E_t$, and $E_z$ respectively. We assume that within the vortex core, radial and tangential velocity components are negligible, and that the fluid is moving vertically upwards, For simplicity we assume, that gradients of, radial and tangential velocity fields, are zero in $z$ direction. The velocity fields are given by (Massey 1968),

$$v_r = \frac{\Lambda}{2\pi r} \tag{1}$$

$$v_t = \frac{\Gamma}{2\pi r} \tag{2}$$

Thermal upcurrent, $v_z$, at the core, can be arrived at by using continuity equation at boundaries of the vortex core,

$$\frac{dv_z(z)}{dz} = \frac{\Lambda}{4\pi r_c^2} \tag{3}$$

which on integration gives,

$$v_z(z) = \frac{\Lambda z}{4\pi r_c^2} \tag{4}$$

We assume that $v_z$ is constant along horizontal direction, but varies with $z$,. Using,

$$v_r(R) = 10 \text{ Kilometers/hour} \tag{5}$$

gives, following relation between suction and effective suction radius,

$$R = 0.0572958\Lambda \tag{6}$$

Before the reactor is switched on , the velocity field of the control volume can be taken to be uniformly zero.

$$v_r(0) = 0 = v_t(0) = v_z(0) \tag{7}$$

As the reactor is switched on, the velocity fields build up, till the point when steady state is reached, and the time derivatives of velocity vectors are zero, due to viscous losses.

$$\dot{v}_r(T_S) = 0 = \dot{v}_t(T_S) = \dot{v}_z(T_S) \tag{8}$$

where, $T_S$, represents the time instant when steady state is reached. In steady state, rate of energy loss due to viscosity and turbulence equals reactor power. Total energy $E_{Total}$, imparted to the atmosphere by the nuclear reactor, operating at constant power $P$, before reaching the steady state, attained after a time interval $\tau$, is given by,

$$E_{Total} = \int_0^\tau P(t)dt = P\tau_s \tag{9}$$

Total kinetic energy $E$ of the air within the control volume $V$,

$$E = E_r + E_t + E_z \tag{10}$$

Applying principle of equipartition of energy, to the radial, tangential, and vertical degrees of freedom, within the control volume, gives the following relation,

$$\frac{E}{3} = E_r = E_t = E_z \tag{11}$$

which, fixes the ratio,

$$\frac{\Gamma}{\Lambda} = 1 \tag{12}$$

Kinetic energy of components of fluid within control volume are arrived at as follows. Integrating over volume external to vortex core gives,

$$E_t = \int_{r_c}^{R} \int_{0}^{2\pi} \int_{0}^{H} \frac{\rho H \Gamma^2}{8\pi^2 r} dr d\phi dz = \frac{\rho H \Gamma^2}{4\pi} \ln \frac{R}{r_c} \tag{13}$$

and,

$$E_r = \int_{r_c}^{R} \int_{0}^{2\pi} \int_{0}^{H} \frac{\rho H \Lambda^2}{8\pi^2 r} dr d\phi dz = \frac{\rho H \Lambda^2}{4\pi} \ln \frac{R}{r_c} \tag{14}$$

where, $\rho$ is air density (assumed constant). Integrating within the vortex core gives,

$$E_z = \int_{0}^{H} \frac{\rho \Lambda^2 z^2}{32\pi r_c^2} dz = \frac{\rho \Lambda^2 H^3}{96\pi r_c^2} \tag{15}$$

The total energy therefore is,

$$E = \frac{\rho H}{4\pi} \left( \Gamma^2 + \Lambda^2 \right) \ln \frac{R}{r_c} + \frac{\rho \Lambda^2 H^3}{96\pi r_c^2} \tag{16}$$

Interestingly, the equipartition of energy, $E_r = E_z$, fixes $r_c$, by the following equation,

$$\ln \frac{R}{r_c} = \frac{H^2}{24 r_c^2} \tag{17}$$

Bernoulli's equation gives the following axi-symmetric pressure field $p$, for region outside vortex core,

$$p(r,z) = p_a(z) - \frac{\rho}{8\pi^2} \left( \Gamma^2 + \Lambda^2 \right) \tag{18}$$

where, $p_a(z)$, is atmospheric pressure.

Since $E = P\tau = 3E_r$, we have another relation between $P$ and $\Lambda$,

$$P = \frac{3\rho H \Lambda^2}{4\pi\tau} \ln \frac{R}{r_c} \tag{19}$$

Assuming nuclear reactor power $P$, is constant, the above equation shows gives the time dependence of suction $\Lambda$, due to energy injected into atmosphere by the nuclear reactor. For brevity, we express above equation as,

$$P = A \left( \ln \frac{R}{r_c} \right) \tag{20}$$

where, using eq. (6),

$$A = \frac{3\rho H \Lambda^2}{4\pi\tau}$$

$$= 2.41917 \frac{R^2}{\tau}, \quad for\ H = 10^4 m, \quad and \quad \rho = 1.293 \quad kg/m^3 \tag{21}$$

See figure 4 for a plot of $P$ versus $R$ , and figure 5, for a 3-D plot of $P\ vs\ R\ vs\ \tau$ (core radius, $r_c=1$ Kilometer in both). Eq. (20), and eq.(6), implicitly give $\Lambda$ ,

$$\Lambda = \left( \frac{4\pi\tau P}{3\rho H \ln(0.0572958\Lambda/r_c)} \right)^{1/2} \tag{22}$$

Numerical solution of eq.(22) in conjunction with eq.(1), give a relation between $v_r$ and $r$.

### 3.2.2 Viscous Energetics

In the steady state, indicated by eq.(8), viscous loss equals ANHISHA's reactors' output. Suction $\Lambda$ , and circulation $\Gamma$ , are at their constant, peak value. Main viscous loss will be the drag exerted by the water (in hurricane scenario), and desert floor sand (in desert scenario). We have,

$$P = -\int_0^{2\pi}\ \int_{r_c}^{\infty} \frac{1}{2}\rho v^3(r) C_D r\,dr\,d\phi$$

$$= \frac{1}{4\pi^2}\rho C_D \Lambda^3 \left( \frac{1}{r_c} \right) \tag{23}$$

$$= \frac{1}{4\pi^2}\rho C_D \left( \frac{R}{0.0572958} \right)^3 \left( \frac{1}{r_c} \right)$$

where, $C_D$ is average value of the drag coefficient (assumed constant), and $v$ is local velocity, given by eq.(1) and (2). Plot for $P\ vs\ R$ for $C_D=10^{-10}$, is given in figure 6.

Velocity field $v(r)$, for a given power $P$, is given by,

$$v(r) = \frac{1}{r} \left( \frac{r_c P}{2\pi\rho C_D} \right)^{1/3} \tag{24}$$

### 3.2.3 Heat Exchanger

Next we consider the heat transfer from reactor to atmosphere. This will help us relate reactor power, to other parameters of ANHISHA's vortex-suction model. We assume the following 'black-box', approach to the heat exchanger. Let $C_E$, be heat transfer coefficient from nuclear reactor to air. Let this transfer occur across a heat exchanger area $A$, which is maintained at a constant temperature $T_E$. Lets assume that the heat exchanger geometry is that of a cylinder of height $z_E$ and radius $r_E$ . Let, $T_0$ and $T_f$ , be the temperatures of air, before and after heating, respectively. Let $\sigma_E$ ,be mass flux into the heat exchanger. Assuming that the heating of air is accomplished at constant pressure, we have following relation for rate of heat transfer to air,

$$\dot{Q} = C_p \sigma_E \left( T_f - T_o \right)$$

where $C_p$ ,is specific heat of air, at constant pressure. Now energy loss by the nuclear reactor is,

$$P = C_E A_E (T_E - T_0)$$

### 3.3 ANHISHA's Hurricane Scenarios

A number of computational models (Zhang et. al. 2000; Challa et. al. 1998; Nguyen et. al. 2001; Zhu et. al. 2001) of hurricanes have been made. There also exist analytical models (Kieu, 2004a; 2004b), and statistical models for predicting hurricane tracks (Hall and Jewson, 2005a; 2005b; 2005c). The existing computational and analytical models of hurricanes, in combination with available atmospheric codes, and observational data, can be used to explore these hurricane scenarios. Technical plan for ANHISHA's hurricane control scenario, is as follows.

(1) Undertake a survey of
    (i) Available atmospheric codes,
    (ii)  Hurricane computational models,
    (iii) Statistical models for predicting hurricane paths,
    (iv) Analytical  models of hurricanes.
    (v) Available observational data on hurricanes.
    (vi) Analytical and computational models for assessing losses due to viscous boundary layers, and turbulence.

(2) Integrating, computational, analytical, and statistical  models of hurricanes, into atmospheric codes. The code resulting as a result of this integration will be referred to the "*Integrated Hurricane Code*" (IHC).

(3) Computer simulations of hurricanes using the integrated hurricane code. Simulation of  real hurricanes which have occurred in past.

(4) Comparison of predictions of integrated hurricane code with existing observational data on hurricanes.

(5)  Conduct numerical experiments, using the Integrated Hurricane code, in which ANHISHA is introduced as heat sources, near the hurricane. Studying time evolution of hurricane parameters, path, intensity, velocity and pressure fields, in response to various combinations of ANHISHA's heat sources. Study effect of varying reactor distance vectors, power, operation time period, on hurricane evolution.

We plan to apply ANHISHA scenarios to the following geographical locales, during time periods of high frequency of occurrence of cyclones and hurricanes –
    (1) Gulf of Mexico, - hurricanes approaching  Florida, US south coast states.
    (2) Western Atlantic, - hurricanes approaching US east coast.
    (3) Western Pacific, - hurricanes approaching California,
    (4) Eastern Pacific, - cyclones and hurricanes approaching Japan, Taiwan, Philippines.
    (5) Southern Pacific – cyclones approaching New Zealand, Australia

### 3.4 ANHISHA's Desert Rain Scenarios

Our plan in this scenario, is to study evolution of humidity field, in response to sustained low pressure region created in desert interior, using ANHISHA. We  wish to use statistical models for calculating  precipitation in desert interior, in response to increased humidity levels. Models of calculating concentration of aerosols in desert air, due to winds generated by ANHISHA, will also be required. Here again we would wish to estimate the effect on physical fields $v,p,$ and $h$, due to variation in $N$, $P_i$, and $D_i$. The atmospheric code will be applied to following geographic locales –

    (1) Sahara,
    (2) Australian interior desert,
    (3) Southern US deserts of Texas, California, Nevada etc..
    (4) Indian Thar desert in Rajasthan.

# 4. Significance

1. The proposed research will integrate available research on hurricane, and improve understanding of hurricane physics. It will pave way for collaboration among experts working in various areas related to hurricane research, - namely, analytical models, computational models, statistical models for path prediction, observational data using remote sensing, and dropsondes.
2. Envisaged studies based upon ANHISHA, will help in understanding of interaction between adjacent low pressure systems.
3. A validation of ANHISHA's desert rain concept, will pave way for a possible future practical implementation, of this idea, with related benefits, outlined in section 2.
4. ANHISHA introduces new peaceful uses of nuclear energy, for creating more genial environments.
5. It will contribute to field of geo-engineering, and help in issues related to modeling such processes, and analyzing their outcomes.

# References

1. Challa, M., Pfeffer, R.L., Zhao, Q., Chang, S.W. (1998): *J. Atmos. Sci.,* Vol. 55, pp. 2201-2219.
2. Emanuel, K., (2006a): physics/0601050, xxx.lanl.gov
3. Emanuel, K. , (2006b): physics/0601051, xxx.lanl.gov
4. Gray, W.M., (2005): *Nature,* Vol. 309, pp. 1844-1846
5. Hall, T. and Jewson, S., (2005a): physics/0512135, xxx.lanl.gov
6. Hall, T. and Jewson, S., (2005b): physics/0512124, xxx.lanl.gov
7. Hall, T. and Jewson, S., (2005c): physics/0512103, xxx.lanl.gov
8. Jieu, C.Q. (2004a): physics/0407073, xxx.lanl.gov
9. Jieu, C.Q. (2004b): physics/0408044, xxx.lanl.gov
10. Kormondy, E.J., (1996): *Concepts of Ecology,* Prentice-Hall Inc., New Jersey.
11. Massey, B.S., (1968): *Fluid Mechanics*, Van Nostrand, London, 1968.
12. Nguyen, M.C., Zhu, H., Smith, R.K., Ulrich, W. (2001): *Q. J. R. Meteo. Soc.,* Vol. 128, pp. 1-20.
13. Zhang, D-L, Liu, Y. and Yau, M.K. (2000): *Mon. Wea. Rev.*, Vol. 128, pp. 3772-3788.

14. Zhu, H., Smith, R.K., Ulrich, W. (2001): *J. Atmos. Sci.,* Vol. 58, pp. 1924-1944.

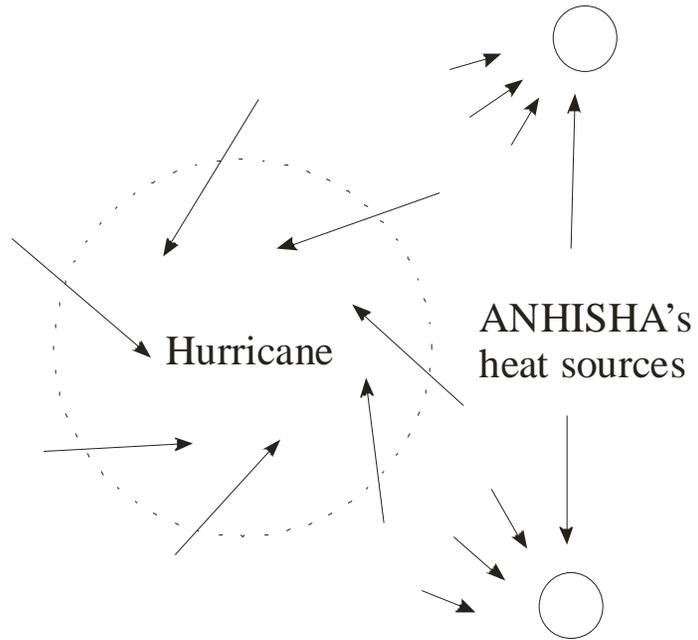

Figure 1
Perturbing hurricane's velocity field with ANHISHA's heat sources. ANHISHA creates wind shear, which can dirupt hurricane's energy feedback mechanism.The artificial low pressure system created by ANHISHA, can recurve hurricane track.

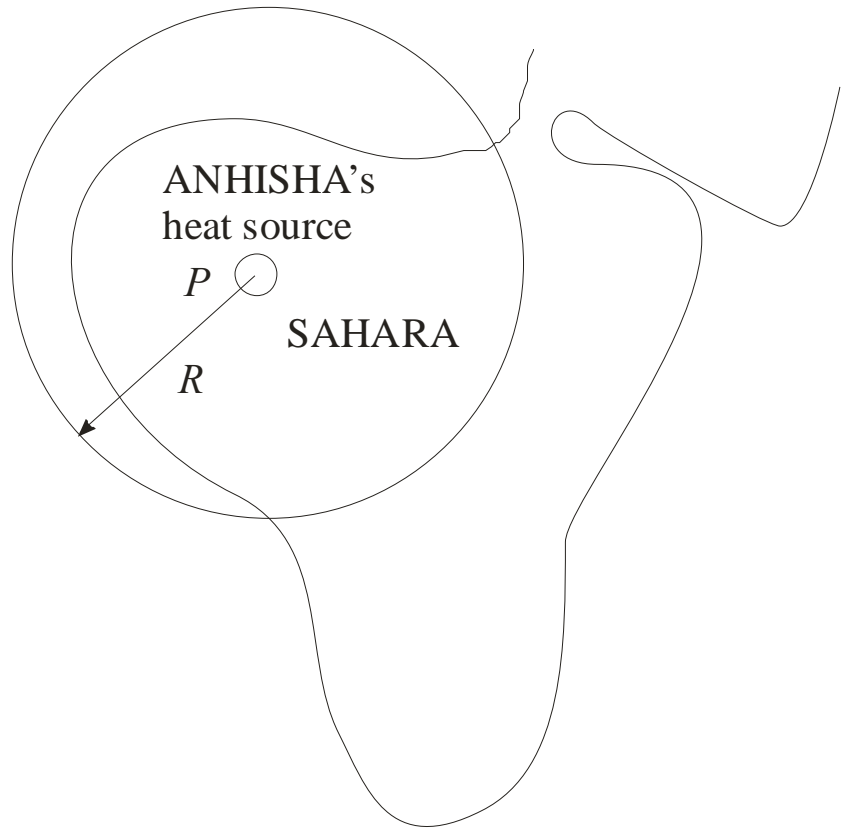

Figure 2
ANHISHA's desert scenario. Low pressure system created by ANHISHA, in Sahara desert, pulls in moist air from Atlantic ocean and Mediteranian sea.

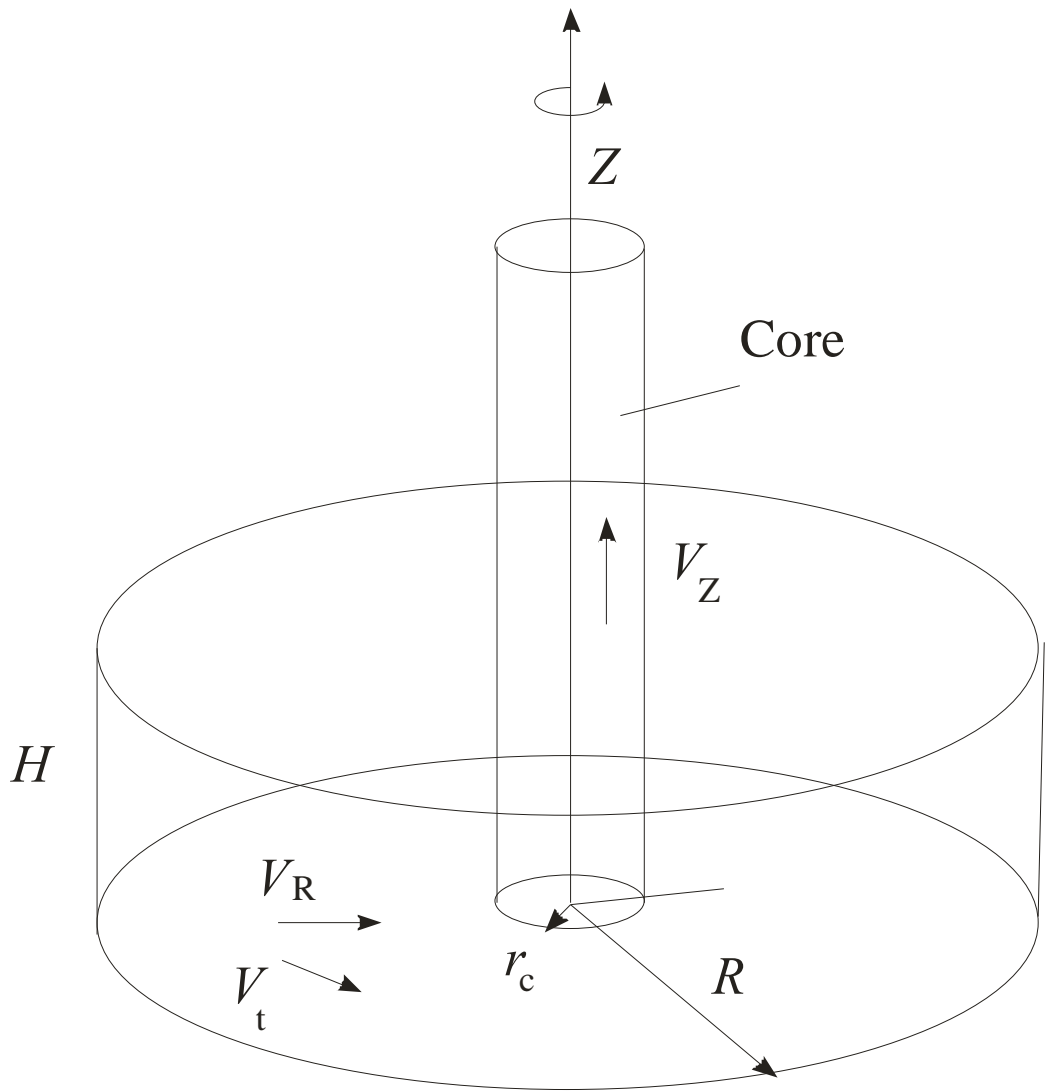

Control Volume

Figure 3
Geometry of ANHISHA's vortex-suction system

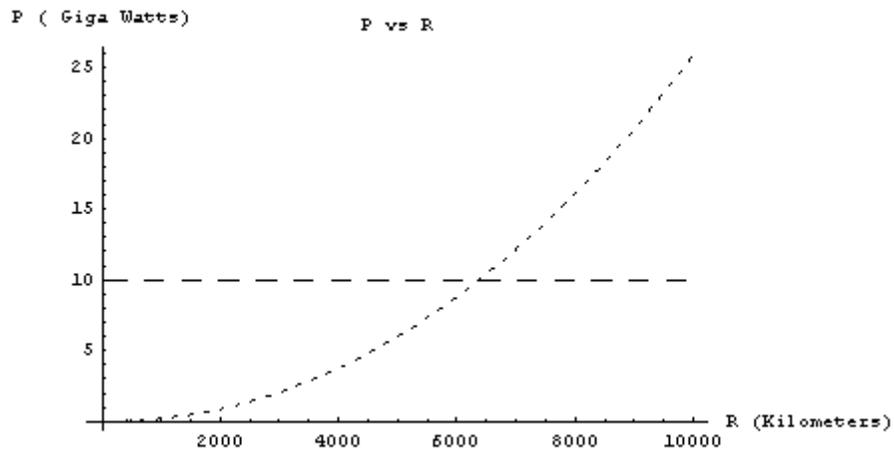

Figure 4

Nuclear Reactor power *P* versus the effective suction radius *R,* for *H~10* Kilometers, *τ~1* day, and *r_c=1* Kilometer. It can be seen, that after a single day of operation of ten 1 Giga Watt reactors, effective suction radius extends to over 6000 Km (in absence of viscous losses). The picture however changes drastically when viscous losses are taken into account, as in figure 6.

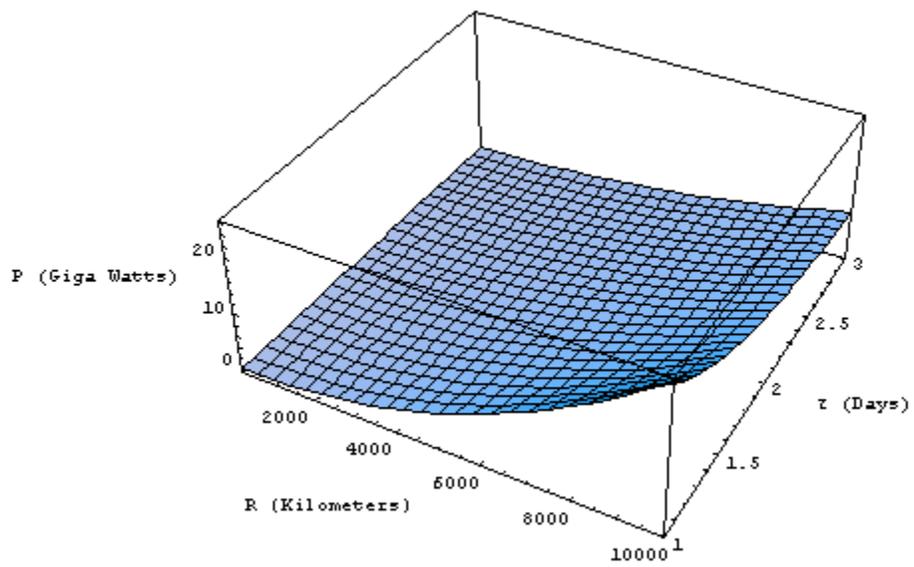

Figure 5
3-D Plot of $P$ vs $R$ vs $\tau$ (inviscid case).

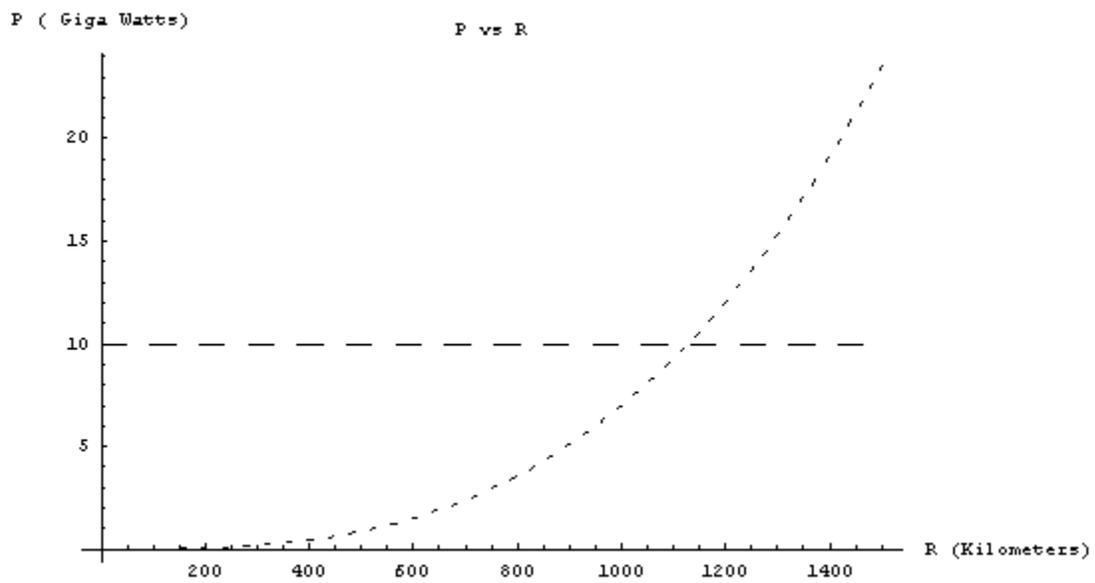

Figure 6

*P* (Reactor Power) vs *R* (effective suction radius), for viscous case, for drag coefficient
$C_D=10^{-10}$, $r_c=10$ Km.

B nmb